\documentclass{ifacconf}

\usepackage{natbib}     

\usepackage{amsmath}    
\usepackage{amssymb}    

\usepackage{graphicx}   
\usepackage{float}      
\usepackage{adjustbox}  
\usepackage[caption=false]{subfig} 

\usepackage{tikz}       
\usepackage[siunitx,RPvoltages]{circuitikz} 
\usepackage[mode=buildnew]{standalone}

\usepackage{color}
\definecolor{JLcolor}{rgb}{1,0.4,0}

\begin{document}
\begin{frontmatter}

\title{Frequency domain parametric estimation \\ of fractional order impedance models \\ for Li-ion batteries\thanksref{footnoteinfo}} 

 \thanks[footnoteinfo]{This research was financially supported by the Research Foundation Flanders (FWO-Vlaanderen, grant nr G.0052.18N) and by the Flemish Government (Methusalem Fund METH1).}

\author[First]{Freja Vandeputte} 
\author[First,Second]{No\"el Hallemans}  
\author[Second]{Jishnu Ayyangatu Kuzhiyil}
\author[Second]{Nessa Fereshteh Saniee} 
\author[Second]{Widanalage Dhammika Widanage} 
\author[First]{John Lataire}

\address[First]{Vrije Universiteit Brussel, Brussels, Belgium \\ (e-mail: Freja.Vandeputte@vub.be)}
\address[Second]{University of Warwick, Coventry, United Kingdom}

\begin{abstract} 
The impedance of a Li-ion battery contains information about its state of charge (SOC), state of health (SOH) and remaining useful life (RUL). 
Commonly, electrochemical impedance spectroscopy (EIS) is used as a nonparametric data-driven technique for estimating this impedance 
from current and voltage measurements.  
In this article, however, we propose a consistent parametric estimation method 
based on a fractional order equivalent circuit model (ECM) of the battery impedance. 
Contrary to the nonparametric impedance estimate, which is only defined at the discrete set of excited frequencies, the parametric estimate can be evaluated in every frequency of the frequency band of interest.
Moreover, we are not limited to a single sine or multisine excitation signal. 
Instead, any persistently exciting signal, like for example a noise excitation signal, will suffice. 
The parametric estimation method is first validated on simulations and then applied to measurements of commercial Samsung 48X cells.
For now, only batteries in rest, i.e. at a constant SOC after relaxation, are considered. 
\end{abstract}



\end{frontmatter}



\section{Introduction}
Li-ion batteries are everywhere.  
They are found in electronic devices, electric vehicles and energy storage systems, 
and thus there is a lot of research being done to improve their lifetime, performance and cost. 
\linebreak 
Electrochemical impedance spectroscopy (EIS) is a powerful non-invasive technique for studying the physical processes occurring in the batteries \citep{EIS}. 
By~measuring the voltage response to a current excitation, the impedance of a battery can be estimated.
In classical EIS, the battery impedance 
is assumed to be both linear \linebreak 
and time-invariant.
In practice, these properties can be realised by measuring in steady state, so after relaxation, with a small amplitude zero mean excitation signal. \linebreak
Indeed, since the relationship between the current and the voltage in a battery, as described by the Butler–Volmer equation, is actually slightly nonlinear, the amplitude of the excitation current must be kept sufficiently small in order to satisfy the linearity assumption.
If, in addition, \linebreak the excitation current is also zero mean, the state of charge (SOC) of the battery will remain approximately constant, 
such that the stationarity assumption is fullfilled as well.
Classical EIS, as implemented in commercial cyclers and potentiostats, estimates the battery impedance  nonparametrically, 
i.e.\ in a set of selected frequencies. 
The estimated impedance can then be interpreted by means of an equivalent electrical circuit \citep{ECM}.    
This analysis relies on the fact that physical \linebreak processes such as migration and accumulation of charge in the battery exhibit the same response as electric resistors and capacitors, respectively.
At low frequencies, the main physical process in a battery is diffusion. 
In an equivalent circuit, diffusion can be represented by a Warburg element, whose impedance is a function of the Warburg variable $\sqrt{s}$, where $s$ is the Laplace variable.
Hence, in the frequency domain, linear time-invariant diffusion results in a transfer function in $\sqrt{s}$. In the time domain, this corresponds to a fractional differential equation (FDE) \citep{FDEPintelon,FractionalModels}. 

The contribution of this article is that we immediately start 
from a Randles equivalent circuit and its corresponding FDE 
to estimate the battery impedance parametrically in the frequency domain. 
Since the equation error of the FDE is linear in the parameters, 
its minimisation boils down to a total least squares (TLS) problem.
%
While an odd random phase multisine is the preferred excitation signal for nonparametric estimation \citep{Multisine}, parametric estimation allows for the use of any persistently exciting signal, like for example Gaussian white noise.
Other possible excitation signals include a sequence of charge and discharge pulses,
as used in pulse power characterisation (PPC) tests, or a pulse-multisine, which consists of an odd random phase multisine superimposed on such a pulse sequence to better mimic drive-cycle characteristics \citep{PulseMultisine1}. 

First, in Section~\ref{sec:sec2}, we introduce the impedance based battery model and derive the expression for the impedance in the Warburg variable $\sqrt{s}$.
In Section~\ref{sec:sec3}, we explain how the current and voltage data is collected. 
Next, in Section~\ref{sec:sec4}, we discuss the fractional differential equation.
The parametric estimation algorithm is described in Section~\ref{sec:sec5}.
Finally, the algorithm is validated on a simulation example in Section~\ref{sec:sec6} and applied to measurements in Section~\ref{sec:sec7}.



\section{Impedance based battery model}
\label{sec:sec2}
\begin{figure} [H]
    \centering
    \includegraphics{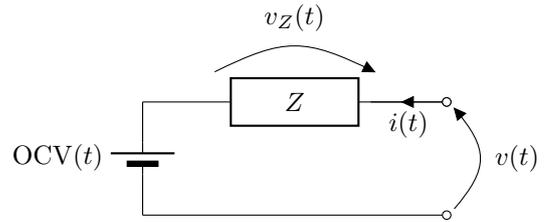}
    \caption{Electrical circuit model of a Li-ion battery. 
    The current through and voltage over the battery are denoted as $i(t)$ and $v(t)$, respectively.}
    \label{fig:BatteryModel}
\end{figure}

A simplified electrical circuit model of a Li-ion battery is shown in Fig.~\ref{fig:BatteryModel}. The battery is modelled as a voltage source, called the open circuit voltage (OCV), in series with an impedance $Z$. 
The voltage $v(t)$ over the battery can then be written as
\vspace{1mm}
\begin{equation}
    v(t) = \mathrm{OCV}(t) + v_Z(t), \\[1mm]
    \label{eq:voltage}
\end{equation}
where both terms depend on the current $i(t)$ through the battery. 
The OCV depends on the SOC of the battery, which in turn depends on the current through Coulomb counting 
\citep{Noel2022}
\begin{equation}
    \mathrm{SOC}(t) = \mathrm{SOC}_0 + \frac{100}{3600C}\int_0^t i(\tau)d\tau, 
    \label{eq:SOC}
\end{equation}
where $\mathrm{SOC}_0$ is the initial SOC in \% and $C$ is the capacity of the battery in Ah.
A 100\% SOC means that the battery is fully charged, while a 0\% SOC means that it is fully discharged.
The voltage $v_Z(t)$ directly depends on the current through the impedance operator $Z$
\vspace{1mm}
\begin{equation}
   v_Z(t) = Z\{i(t)\}. \\[1mm]
\end{equation}
If the battery impedance $Z$ is assumed to be 
a linear time-invariant (LTI) system, 
the voltage over the impedance can easily be computed via a multiplication in the frequency domain as follows
\vspace{1mm}
\begin{equation}
    v_Z(t) = \mathcal{L}^{-1}\{Z(s)I(s)\}, \\[1mm]
    \label{eq:out_LTI}
\end{equation}
where $\mathcal{L}^{-1}$ denotes the inverse Laplace transform, 
$Z(s)$ is the impedance 
and $I(s)$ is the Laplace transform of $i(t)$, 
\begin{equation}
   I(s) = \mathcal{L}\{i(t)\} = \int_0^{\infty}i(t)e^{-st}dt.
\end{equation}
%
In reality, the relation between the current and the voltage in a battery is inherently slightly nonlinear, and thus nonlinear distortions will be introduced. 
If the root mean square (RMS) value of the excitation current
is sufficiently small however, 
the nonlinear distortions will not be too strong, such that the use of the linear model \eqref{eq:out_LTI} is justified. 

Time-invariance is achieved by keeping the SOC approximately constant after relaxation. 
Hence, it follows from~\eqref{eq:SOC} that the current $i(t)$ should be zero mean with a small 
RMS value. 
As the SOC is constant, the OCV is constant as well, such that 
around this operating point,
\eqref{eq:voltage} becomes 

\begin{figure} [H]
    \centering
    \includegraphics[scale = 0.75]{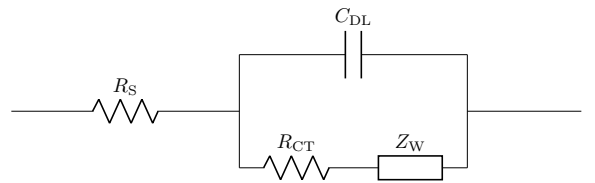}
    \caption{Randles ECM for the battery impedance $Z$ with a Warburg element to model the diffusion.}
    \label{fig:ECM}
\end{figure}

\begin{figure} [H]
    \centering
    \includegraphics[scale = 0.75]{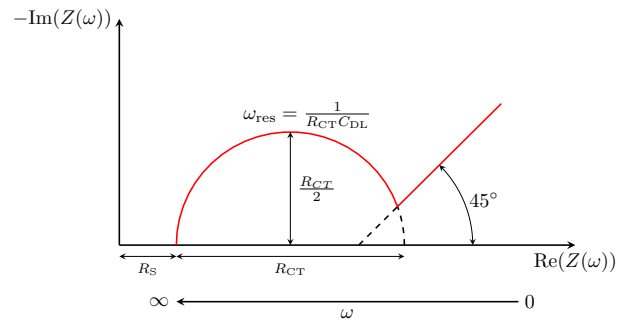}
    \vspace{-2mm}
    \caption{Nyquist plot of the Randles ECM in Fig.~\ref{fig:ECM}. }
    \label{fig:Nyquist}
\end{figure}

\vspace{-3mm}
\begin{equation}{}
    v(t) = \mathrm{OCV} + \mathcal{L}^{-1}\{Z(s)I(s)\}. \\[1.5mm]
    \label{eq:voltage_LTI}
\end{equation}
%
The battery impedance $Z(s)$ can be modelled parametrically by an equivalent circuit model (ECM).
The equivalent circuit components do not necessarily have physical meaning, but they give information about the electrochemical behaviour of the battery. 
The Randles ECM in Fig.~\ref{fig:ECM} 
consists of an electrolyte resistance~$R_\mathrm{S}$, 
a double layer capacitance~$C_\mathrm{DL}$, a~charge transfer resistance~$R_\mathrm{CT}$ and a Warburg diffusion element~$Z_{\mathrm{W}}$. 
In the Nyquist plot in Fig.~\ref{fig:Nyquist}, 
the electrolyte resistance $R_{\mathrm{S}}$ can be found at the real axis intercept, as it is the only remaining component at high frequencies, when the capacitance $C_\mathrm{DL}$ acts as a short. 
The parallel connection $C_\mathrm{DL}//R_\mathrm{CT}$ 
results in a semicircle with diameter $R_\mathrm{CT}$ that reaches a maximum at the resonance frequency 
\begin{equation}
    \omega_{\mathrm{res}} = \frac{1}{R_\mathrm{CT}C_\mathrm{DL}}.
\end{equation}
The Warburg element models semi-infinite linear diffusion in the battery, i.e.\ diffusion from the electrolyte to a large planar electrode.
The Warburg impedance 
is given by
\begin{equation}
    Z_\mathrm{W}(\omega) = (1-j)\frac{\sigma}{\sqrt{\omega}},
\end{equation}
which can be rewritten as the impedance of a constant phase element 
with a constant phase of $-45^{\circ}$
\begin{equation}
    Z_\mathrm{W}(\omega) = \frac{\sigma\sqrt{2}}{\sqrt{\omega}} e^{-j\frac{\pi}{4}}
    = \frac{\sigma\sqrt{2}}{\sqrt{j\omega}} .
    \label{eq:ZW_freq}
\end{equation}
Hence, at low frequencies, the Warburg impedance leads to a straight line with a $45^{\circ}$ slope in the Nyquist plot.
\vspace{-2.5mm}
\vspace{-1.1mm}
Substituting $j\omega$ by the Laplace variable $s$ in \eqref{eq:ZW_freq} gives
\begin{equation}
     Z_\mathrm{W}(s) 
    = \frac{\sigma\sqrt{2}}{\sqrt{s}},
\end{equation}
where $\sqrt{s}$ is called the Warburg variable.
Using the Randles ECM in Fig.~\ref{fig:ECM}, 
the total battery impedance in the Laplace variable $s$ becomes
\begin{align}
    Z(s) 
    &= R_{\mathrm{S}} + \frac{1}{\frac{1}{R_\mathrm{CT}+\frac{\sigma\sqrt{2}}{\sqrt{s}}}+sC_\mathrm{DL}} \label{eq:Z_Laplace},
\end{align}

\vspace{-1.5mm} 
which can be rewritten as a rational function in the Warburg variable $\sqrt{s}$
\begin{equation}
    Z(\sqrt{s}) = \frac{b_0+b_1\sqrt{s}+b_2(\sqrt{s})^2+b_3(\sqrt{s})^3}{a_1\sqrt{s}+a_2(\sqrt{s})^2+a_3(\sqrt{s})^3},
    \label{eq:Z_Warburg}
\end{equation}

\vspace{-0.5mm} 
with coefficient $a_1 = 1$ and 
\begin{subequations}
\begin{align}
    &a_2 = \sigma\sqrt{2}C_\mathrm{DL}    &&a_3 = R_\mathrm{CT}C_\mathrm{DL}\\  
    &b_0 = \sigma\sqrt{2}          &&b_1 = R_{\mathrm{S}}+R_\mathrm{CT}\\
    &b_2 = R_{\mathrm{S}}\sigma\sqrt{2}C_\mathrm{DL} &&b_3 = R_{\mathrm{S}}R_\mathrm{CT}C_\mathrm{DL}. 
\end{align}
\label{eq:Coeff}
\end{subequations}

\vspace{-4mm}
Hence, given the coefficients $a_n$ and $b_n$, the equivalent circuit components can be found by solving a nonlinear least squares problem of 6 equations in 4 unknowns. 
\section{Collecting data}
\label{sec:sec3}

The parametric estimation algorithm has the advantage that any persistently exciting signal can be used to excite the system.
In the frequency domain approach of system identification, it is common to apply a multisine as an excitation signal
\begin{equation}
    i(t) = \sum_{k\in\mathbb{H}_\mathrm{exc}} \alpha_{k} \sin{\left(\frac{2\pi k}{T_p} t+\varphi_{k}\right)},
    \label{eq:multisine}
\end{equation}
where $T_p$ is the period of the multisine and $\mathbb{H}_\mathrm{exc}$ is the set of excited harmonics.
Specifically, an odd random phase multisine will be used. 
This entails that $\mathbb{H}_\mathrm{exc}$ is~chosen such that only odd harmonics are excited, i.e.\ $\mathbb{H}_\mathrm{exc}\subset2\mathbb{N}+1$.
Moreover, to be able to cover a large frequency band of multiple decades, the excited harmonics
are quasi-logarithmically distributed. 
The amplitudes $\alpha_{k}$ are user-defined and the random phases $\varphi_{k}$ are uniformly distributed in $[0,2\pi)$. 
Furthermore, we will also apply a zero mean Gaussian white noise signal as an excitation. 
The obtained excitation signal is scaled such that it has the desired RMS value 

\vspace{-4mm}
\begin{equation}
    \tilde i(t) = \frac{\mathrm{RMS}_{\mathrm{des}}}{\mathrm{RMS}\{i(t)\}}i(t). 
\end{equation}

\vspace{-1mm}
In practice, measured signals are sampled and windowed.
In order to avoid aliasing, the sampling rate $f_s$ 
must satisfy the Nyquist criterion, 
whereas leakage will not be present if the measurement window $T$ 
is an integer number of periods of the multisine, i.e.\ $T=PT_p$.
Hence, by measuring $P$ periods of the multisine, the frequency resolution of the measurement $f_{\mathrm{res}} = 1/T$ is higher than the frequency resolution of the multisine $f_0 = 1/T_p = P/T= P f_{\mathrm{res}}$. 

For an LTI impedance, in the case of a multisine excitation
both the current and the voltage spectrum 
will only have contributions at the excited frequencies, though the voltage spectrum will have an additional contribution at DC due to the constant OCV in \eqref{eq:voltage_LTI} \citep{Noel2022}.
The nonparametric estimate of the battery impedance at the nonzero excited frequencies
is then obtained by a simple division of the discrete Fourier transform (DFT)  
spectra of the voltage and the current 
\begin{equation}
    \hat Z(\omega_{k}) = \frac{V(k)}{I(k)} \hspace{4mm} k\in\mathbb{K}_{\mathrm{exc}}=P\mathbb{H}_{\mathrm{exc}},
    \label{eq:Znonpar}
\end{equation}
where $\omega_k = 2\pi k/T$.
The DFT 
of a windowed and sampled signal $x(n) = x(nT_s)$ 
with $T_s=1/f_s$ the sampling period and $N=T/T_s$ the number of samples, is defined as
\begin{equation}
    X(k) = \mathrm{DFT}\{x(n)\} = \frac{1}{N} \sum_{n=0}^{N-1}x(n)e^{-j2\pi kn/N}.
\end{equation}


\section{Fractional differential equation}
\label{sec:sec4}

In the frequency domain, diffusion in the battery results in a rational function in $\sqrt{s}$ \eqref{eq:Z_Warburg}. In the time domain, this corresponds to a so-called fractional differential equation (FDE)
\begin{equation}
    \sum_{n=1}^{N_a} a_n \frac{\mathrm d^{\frac{n}{2}}v(t)}{\mathrm d t^{\frac{n}{2}}} = \sum_{n=0}^{N_b} b_n \frac{\mathrm d^{\frac{n}{2}}i(t)}{\mathrm d t^{\frac{n}{2}}},
    \label{eq:FDE}
\end{equation}
where the Riemann–Liouville fractional derivative of order $1/2$ is defined as
\begin{equation}
     \frac{\mathrm d^{\frac{1}{2}}f(t)}{\mathrm d t^{\frac{1}{2}}} = \frac{1}{\sqrt \pi} \frac{\mathrm d}{\mathrm dt} \int_0^t \frac{f(\tau)}{\sqrt {t-\tau}}d\tau,
\end{equation}
and the fractional derivative of order $n+1/2$ for $n\in\mathbb{N}$ is obtained as
\begin{equation}
     \frac{\mathrm d^{n+\frac{1}{2}}f(t)}{\mathrm d t^{n+\frac{1}{2}}} = \frac{\mathrm d^{n}}{\mathrm d t^{n}}
     \Big(\frac{\mathrm d^{\frac{1}{2}}f(t)}{\mathrm d t^{\frac{1}{2}}}\Big). 
\end{equation}

As a measured
signal $x(t)$  (=$i(t),v(t)$) is in practice windowed and sampled, the effect of these two operations needs to be taken into
account.
Multiplying both sides of \eqref{eq:FDE} with a rectangular window,

\vspace{-3mm}
\begin{equation}
    w(t) = \left\{
    \begin{array}{ll}
        1 & \; t\in[0,T] \\
        0 & \; \text{otherwise} ,
    \end{array}
    \right.
\end{equation}

\vspace{-1mm}
does not alter the FDE. 
Using integration by parts, it can be proven that the Laplace transform of the windowed derivative of order~$n\in\mathbb{N}$ of a signal $x(t)$ is given by \citep{SysIdBookPintelon}
\begin{align}
    &\mathcal{L} \big\{ w(t) \frac{\mathrm d^nx(t)}{\mathrm d t^n}\big\} 
    = \int_0^T\frac{\mathrm d^nx(t)}{\mathrm d t^n}e^{-st}dt \\
    &= s^nX(s) + \underbrace{\sum_{r=0}^{n-1}s^r\big (x^{(n-1-r)}(T)e^{-sT}-x^{(n-1-r)}(0)\big)}_{=R_w(s)}, \nonumber
\end{align}
with $X(s)$ the Laplace transform of the windowed signal 
\begin{equation}
    X(s) = \mathcal{L}\{w(t)x(t)\} = \int_0^Tx(t)e^{-st}dt,
\end{equation}
and $x^{(n)}(t)$ the derivative of order $n$ of the signal 
\begin{equation}
    x^{(n)}(t) = \frac{\mathrm d^nx(t)}{\mathrm d t^n}.
\end{equation}

\pagebreak
The transient polynomial $R_w(s)$ is a polynomial in $s$ of degree $n-1$ that models the effect of the begin and the end conditions. 
Analogously, it can be shown that 
the Laplace transform of the windowed derivative of order $n+1/2$ for $n\in\mathbb{N}$ of a signal $x(t)$ is given by 

\begin{align}
    &\mathcal{L} \left\{ w(t) \frac{\mathrm d^{n+\frac{1}{2}} x(t)}{\mathrm d t^{n+\frac{1}{2}}}\right\} 
    = \int_0^T\frac{\mathrm d^{n+\frac{1}{2}} x(t)}{\mathrm d t^{n+\frac{1}{2}}}e^{-st}dt \\
    &= s^{n+\frac{1}{2}} X(s) + \underbrace{\sum_{r=0}^{n}s^r\big (x^{({n-\frac{1}{2}}-r)}(T)e^{-sT}-x^{({n-\frac{1}{2}}-r)}(0)\big)}_{=R_w(s)} \nonumber
\end{align}


where the transient polynomial $R_w(s)$ is again a polynomial in $s$ of degree $n$.  
If $x(t)$ is band-limited, the Laplace transform of the continuous-time signal $x(t)$ can be approximated by the DFT of the sampled signal $x(n)$ 
\begin{equation}
    s^{\frac{n}{2}}X(s)|_{s = j\omega_k} = (j\omega_k)^{\frac{n}{2}}X(k) + R_a(\sqrt{j\omega_k}),
\end{equation}
where $R_a(\sqrt{j\omega_k})$ is a polynomial to model the alias error. Hence, the errors introduced by sampling and windowing will be captured by an additional polynomial $R(\sqrt{j\omega_k}) = R_w(j\omega_k)+R_a(\sqrt{j\omega_k})$ to be estimated.
This is especially important when using noise as excitation signal.



\begin{figure*}
    \includegraphics[scale = 0.78]{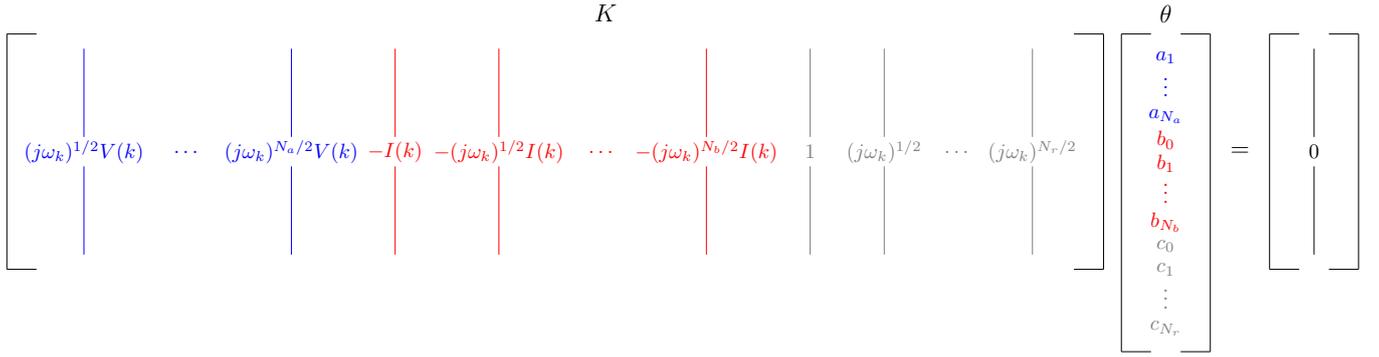} 
    \caption{Graphical representation of the regression matrix $K$ and the parameters to be estimated $\theta$.}
    \label{fig:RegLTI}
\end{figure*}

\section{Parametric impedance estimation}
\label{sec:sec5}
\subsection{Minimising the equation error}
To estimate the parameters $a_n$ and $b_n$, we minimise the equation error, which is of the form
\begin{equation}
    e(t) = \sum_{n=1}^{N_a}a_n \frac{\mathrm d^{\frac{n}{2}}v(t)}{\mathrm d t^{\frac{n}{2}}} - \sum_{n=0}^{N_b}b_n \frac{\mathrm d^{\frac{n}{2}}i(t)}{\mathrm d t^{\frac{n}{2}}}.
\end{equation}
The DFT of this error becomes \citep{FDEPintelon}
\small
\begin{equation}
\begin{aligned}
    E(k) =  \sum_{n=1}^{N_a}a_{n} (j\omega_k)^{\frac{n}{2}}V(k)
    &- \sum_{n=0}^{N_b}b_{n} (j\omega_k)^{\frac{n}{2}}I(k)
     + R(\sqrt{j\omega_k}), 
    \label{eq:DFTeqerr}
\end{aligned}
\end{equation}
\normalsize
where $\omega_k = \frac{2\pi k}{T}$ for each bin $k$ in the user defined frequency window
\begin{equation}
    \mathbb{K}_w = [k_{\mathrm{min}},\dots,k_{\mathrm{max}}],
\end{equation}
to which the estimation is restricted.
$R(\sqrt{j\omega_k})$ is an additional polynomial to capture the effects of windowing and sampling. It is approximated by a linear combination of monomials 
\begin{equation}
    R(\sqrt{j\omega_k}) \approx \sum_{r=0}^{N_r}c_r (j\omega_k)^{\frac{r}{2}}
    \label{eq:RRRR}
\end{equation}
Equation \eqref{eq:DFTeqerr} is linear in the parameters 
\begin{equation}
    \theta = \begin{bmatrix} \cdots &a_{n}& \cdots& b_{n}& \cdots &c_r&\cdots\end{bmatrix} ^\top,
\end{equation}
such that it can be written in matrix form as 
\begin{equation}
    E = K\theta.
    \label{eq:errrr}
\end{equation}
The columns of the regression matrix $K$ are constructed as in Fig.~\ref{fig:RegLTI}. 
%
To ensure that the estimated parameters are real numbers,  
the matrix $K$ is then split in real and imaginary parts
\begin{equation}
    K' = \begin{bmatrix} \mathrm{Re}(K) \\\mathrm{Im}(K) \end{bmatrix}.
\end{equation}
Consider the economy size singular value decomposition (SVD) of this extended regression matrix $K'$, 
\begin{equation}
    K' = U\Sigma W^H.
\end{equation}
The total least squares (TLS) estimate of the parameters, i.e.\ the non-trivial estimate that minimises the equation error,
\begin{equation}
    \hat \theta = \arg\min_{\theta} \sum_{k\in\mathbb{K}_w} |E(k)|^2 \hspace{5mm} \text{ s.t. } ||\hat\theta||_2 = 1,
\end{equation}
corresponds to the last column of $W$. 
The estimated parameters are scaled such that the first parameter $a_1=1$.
The impedance estimate is then obtained by evaluating \eqref{eq:Z_Warburg} with the estimated parameters $\hat a_n$ and $\hat b_n$ in $s=j\omega_k$.


\subsection{Consistent estimation}
Consistency of an estimator means that if the amount of data asymptotically grows to infinity, the estimated parameters  converge to the true parameter values. In other words, if the number of data points increases, the uncertainty on the estimated parameters decreases. Therefore, when constructing an estimator, consistency is a very desirable property. 
The weighted total least squares (WTLS) estimate
%
\begin{equation}
    \hat \theta = \arg\min_{\theta} \sum_{k\in\mathbb{K}_w} \frac{|E(k)|^2}{\sigma^2_E(k)} \hspace{5mm} \text{ s.t. } ||\hat\theta||_2 = 1,
\end{equation}
where $\sigma^2_E(k)$ are the variances of the equation error, 
can be shown to be consistent. The estimated parameters can be found from the thin SVD of the new regression matrix, which is obtained by scaling the rows of the old regression matrix with $1/\sigma_E(k)$ . 
This is an iterative procedure, since to calculate the new parameter estimates, one needs the old parameter estimates to compute 
$\sigma^2_E(k)$, i.e.
\begin{equation}
    \hat \theta_{i+1} = \arg\min_{\theta}  \sum_{k\in\mathbb{K}_w} \frac{|E(k)|^2}{\sigma^2_E(k,\theta_i)} \hspace{5mm} \text{ s.t. } ||\hat\theta_{i+1}||_2 = 1.
\end{equation}
The variances of the equation error can be computed as follows. Equation \eqref{eq:DFTeqerr} is also linear in the spectra, since it can be rewritten as 
\small
\begin{equation}
\begin{aligned}
    E(k) =  \sum_{n=0}^{N_a}\underbrace{a_{n} D_{(j\omega_k)^{n}}}_{\mathcal{A}_{n}} V(k) 
    &- \sum_{n=0}^{N_b}\underbrace{b_{n} D_{(j\omega_k)^{n}}}_{\mathcal{B}_{n}}  I(k)
    + R({\sqrt{j\omega_k}}),
    \label{eq:notcom}
\end{aligned}
\end{equation}
\normalsize
where $D_{(j\omega_k)^{n}}$ is a diagonal matrix with  on the diagonal  $(j\omega_k)^{n}$ with $k = [0,1,\dots,N/2,-N/2,\dots,-1]$. In compact notation, \eqref{eq:notcom} becomes 
\begin{equation}
E = \mathcal{A}V - \mathcal{B}I + R.
\label{eq:eq_error}
\end{equation}
The covariance matrix of the equation error \eqref{eq:eq_error} is then given by
\begin{equation}
C_E  = 
\mathcal{A}C_V\mathcal{A}^{H} + \mathcal{B}C_I\mathcal{B}^{H} -\mathcal{A}C_{VI}\mathcal{B}^{H} - \mathcal{B}C_{VI}^{H}\mathcal{A}^{H},
\end{equation}
where $C_I, C_V$ and $C_{VI}$ are the covariance matrices of the current and voltage spectra.
For an LTI system,  $C_E$ is a diagonal matrix with the variances of the equation error $\sigma^2_E(k)$ on the diagonal.

For an Errors-In-Variables framework, where the input and output signals are disturbed by zero mean additive circularly complex Gaussian noise, the noise distribution is completely described by the noise covariances 
\vspace{1mm}
\begin{subequations}
\begin{align}
C_I(k,k')  &= \mathbb{E}\{N_I(k)N_I^*(k')\} = \delta_{kk'}\sigma_I^2(k) \\
C_V(k,k')  &= \mathbb{E}\{N_V(k)N_V^*(k')\} = \delta_{kk'}\sigma_V^2(k)  \\
C_{VI} (k,k')  &= \mathbb{E}\{N_V(k)N_I^*(k')\} = \delta_{kk'}\sigma_{VI}^2(k). 
\end{align}
\end{subequations}

\vspace{-1mm}
By measuring $P$ periods of the excitation signal, these covariances can be calculated as
\begin{subequations}
\begin{align}
    \sigma_I^2(k) &= {\frac{1}{P\!-1\!}} \sum_{p=1}^{P} (I_p(k)\!-\!\bar{I}(k)\!)(I_p(k)\!-\!\bar{I}(k)\!){^*}\\
    \sigma_V^2(k) &= {\frac{1}{P\!-1\!}} \sum_{p=1}^{P} (V_p(k)\!-\!\bar{V}(k)\!)(V_p(k)\!-\!\bar{V}(k)\!){^*}\\
    \sigma_{VI}^2(k) &=  {\frac{1}{P\!-1\!}} \sum_{p=1}^{P} (V_p(k)\!-\!\bar{V}(k)\!)(I_p(k)\!-\!\bar{I}(k)\!)^*,
\end{align}
\end{subequations}
where $\bar{I}$ and $\bar{V}$ denote the averages of the current and voltage spectra along the $P$ periods, $I_p$ and $V_p$ are the spectra of each period and $^*$ is the complex conjugate.

\section{Simulation}
\label{sec:sec6}
\begin{figure} [H]
    \centering
    \includegraphics[width = 0.5\textwidth]{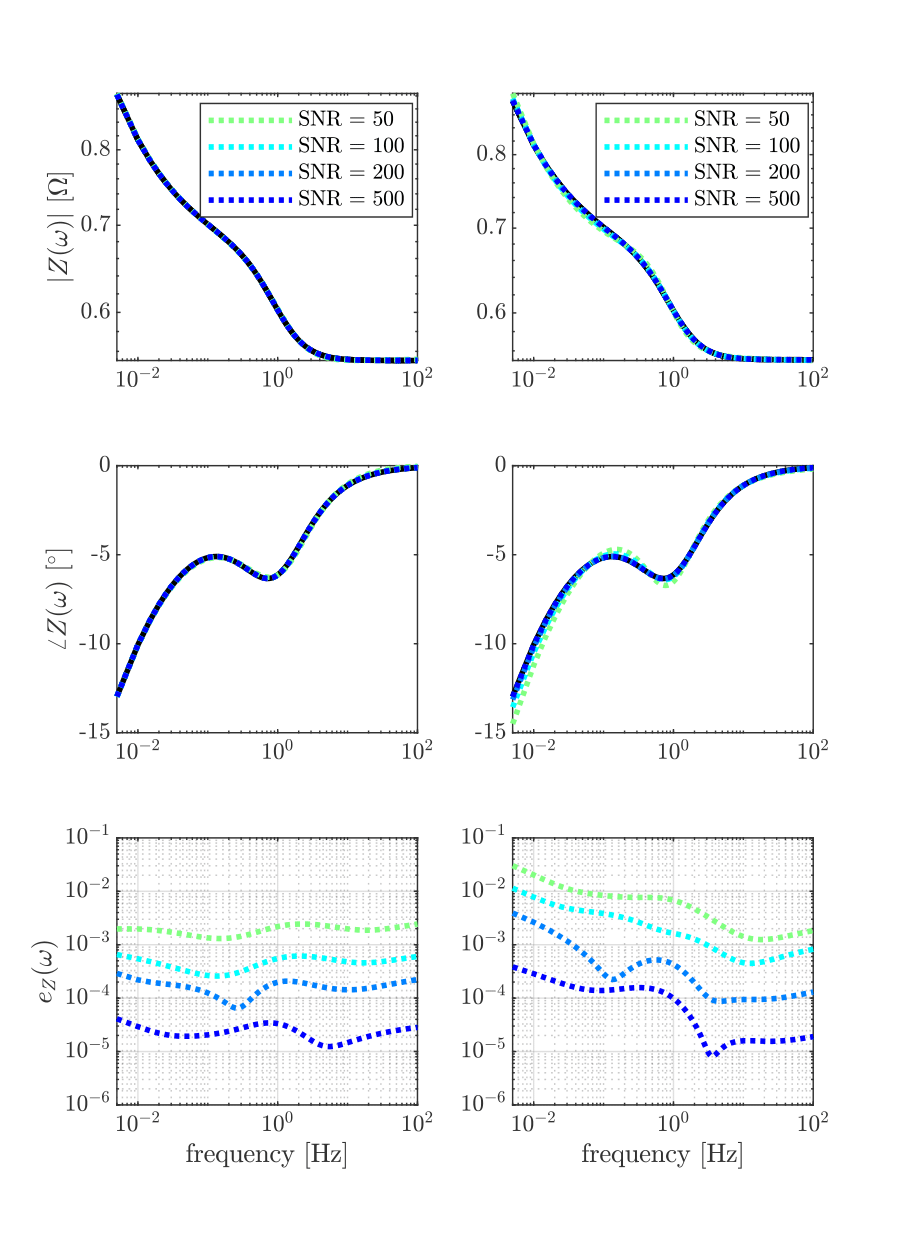} 
    \caption{Bode plot of the simulated impedance \vspace{-0.2mm} $Z_0(\omega)$ (full black line) \vspace{-0.3mm} and the parametrically estimated impedance $\hat Z(\omega)$ (coloured dashed lines) for different SNRs. The relative error $e_Z(\omega)$ remains approximately constant over the frequency for a multisine excitation (left), but decreases with the frequency for a noise excitation (right).}
    \label{fig:SimFreqError_SNR}
\end{figure}

The Randles ECM of the battery impedance is simulated 
in the frequency domain
with $R_{\mathrm{S}} = 551$~m$\Omega$, $R_\mathrm{CT}=~119$~m$\Omega$, $C_\mathrm{DL} = 1464$~mF and $\sigma = 0.0346$~$\Omega/\sqrt{\mathrm s}$ \citep{ECM_componentvalues}. 
Both an odd random phase multisine and zero mean white Gaussian noise with an RMS value of 0.5 are used as an excitation current. 
$P = 5$ periods of $T_p=200$~s are simulated with a sampling rate $f_s=200$~Hz, such that there are $N = 200\,000$ data points.
The current and voltage signals are perturbed by zero mean additive white Gaussian noise, i.e. 
\begin{equation}
   x(t) = x_0(t)+n_x(t) , \\[0.5mm]  
\end{equation}
where $x(t)$ (=$i(t),v(t)$) denotes the noisy signal, $x_0(t)$ is the noiseless signal and $n_x(t)\sim\mathcal{N}(0,\sigma_{n_x}^{2})$ is the noise.  
The Signal-to-Noise Ratio (SNR) is then given by
\vspace{1mm}
\begin{equation}
   \mathrm{SNR}\{x(t)\} = \frac{\mathrm{RMS}\{x_0(t)\}}{\sigma_{n_x}}.{}
\end{equation}
Fig.~\ref{fig:SimFreqError_SNR} shows the Bode plot of the simulated battery impedance $Z_0(\omega)$, \vspace{-0.5mm} obtained by evaluating \eqref{eq:Z_Laplace} in $s=j\omega$, and the parametrically estimated impedance $\hat Z(\omega)$  for different SNRs. The estimated transient polynomial has order $N_r=1$ and the number of iterations for consistent estimation is 10. 
The relative error over the frequency, 
\begin{equation}
   e_Z(\omega) = \frac{|Z_0(\omega)-\hat Z(\omega)|}{|Z_0(\omega)|},
\end{equation}

\pagebreak
decreases for increasing SNR.
For a multisine excitation, the simulated and estimated impedance coincide almost perfectly (relative error of less than 0.3\% for an SNR of~50), while for a noise excitation, \vspace{-0.5mm} 
there is a significant discrepancy between $Z_0(\omega)$ and $\hat Z(\omega)$ at the low frequencies for the smaller SNRs (relative error decreases with the frequency from 3\% to less than 0.3\% for an SNR of 50).   









\pagebreak
\section{Measurements}
\label{sec:sec7}
\begin{figure} [H]
    \centering
    \includegraphics[width = 0.5\textwidth]{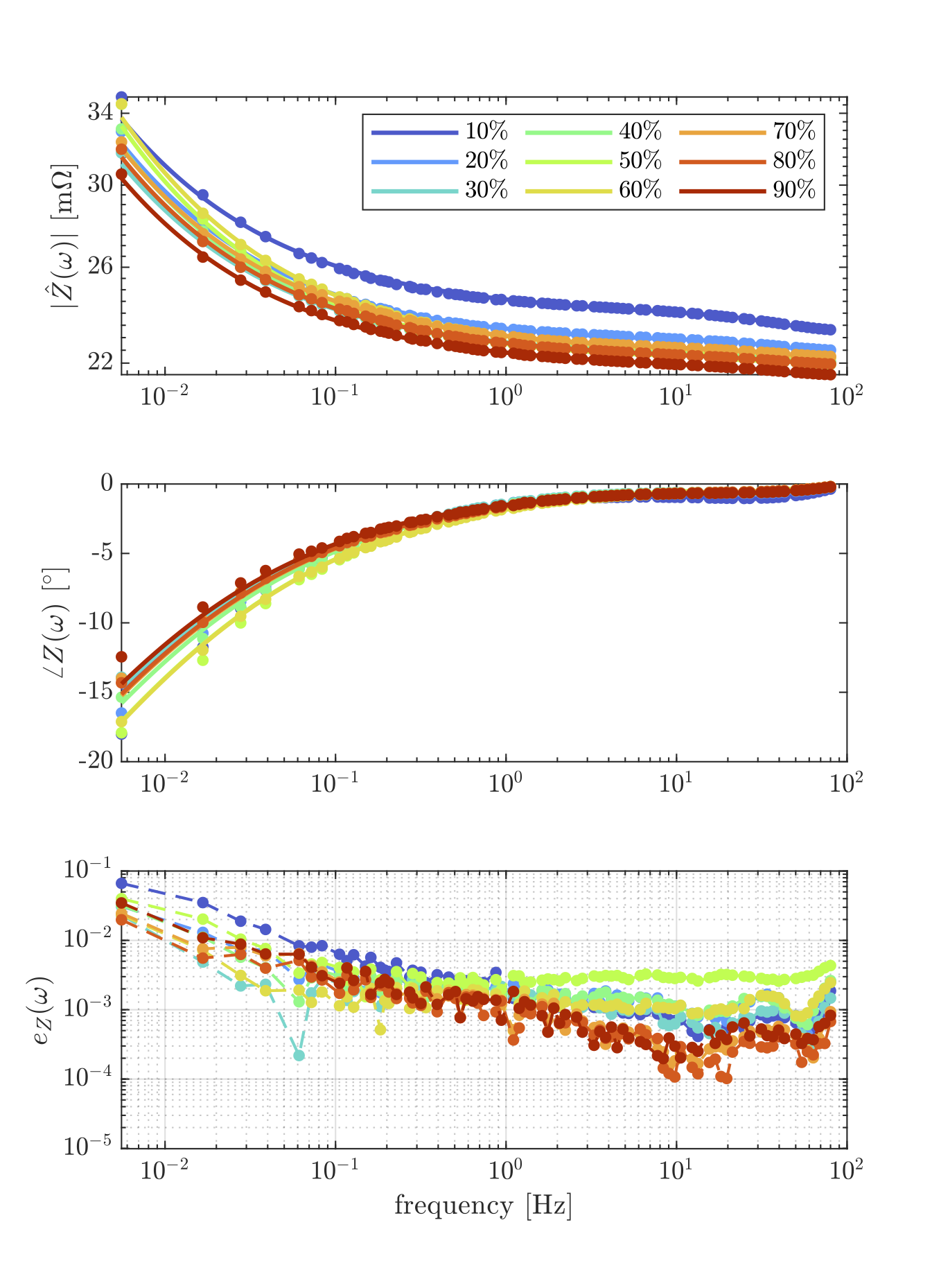}
    \caption{Bode plot of the nonparametric (discrete dots) and parametric (continuous lines) impedance estimates at different SOC levels. The relative error $e_Z(\omega)$ between the nonparametric and parametric estimates is larger at the low frequencies.}
    \label{fig:Data45X}
\end{figure}

EIS measurements were performed using a Gamry Interface 5000 potentiostat on a pristine, commercially available Samsung 48X INR21700 Li-ion battery with a 
$\mathrm{LiNiCoAlO_2}$ (NCA)
cathode and a Si-Gr anode. 
The measurements were done after relaxation at 9 different SOC levels, going from 10\% to 90\% in steps of 10\%.
The excitation current was an odd random phase multisine with an RMS value of 0.5 and a period of $T_p =180$~s. 
The 76 excited frequencies were quasi-logarithmically distributed within the band [5.6 mHz, 80 Hz].
The current and voltage signals were measured for $P = 10$ periods of the multisine,
i.e.\ for $T = 1800$~s, at a sampling rate $f_s = 200$~Hz, such that there were $N = 360\,000$ data points. The order of the estimated transient polynomial has order $N_r=1$ and the number of iterations for consistent estimation is 10.
The Bode plot of the obtained nonparametric \eqref{eq:Znonpar} and parametric impedance estimates is shown in Fig.~\ref{fig:Data45X}. The relative error between them
\begin{equation}
   e_Z(\omega) = \frac{|\hat Z_{\mathrm{nonpar}}(\omega) - \hat Z_{\mathrm{par}}(\omega)|}{|\hat Z_{\mathrm{nonpar}}(\omega)|},
\end{equation}
decreases from less than 10\% at the lower excited frequencies to less than 1\% at frequencies higher than 100 mHz.



\vspace{-2.5mm}

\pagebreak
\section{Conclusion}
\label{sec:sec8}
An algorithm for parametrically estimating the linear time-invariant impedance of a Li-ion battery from current and voltage measurements is successfully implemented. 
The underlying parametric fractional order model is a Randles equivalent circuit
model 
with a Warburg element to model diffusion. 
The equation error of the corresponding FDE, computed in the frequency domain, is linear in the parameters, such that its minimisation becomes a TLS \linebreak estimation problem, which can be solved with the SVD of the regression matrix.
Weighting the regression matrix with the variances of the equation error makes the \linebreak estimation consistent.
While the nonparametric impedance estimate is only defined at the discrete set of excited frequencies, the parametric estimate can be evaluated in every frequency of the frequency band of interest. 
Moreover, the parametric estimation algorithm works for any persistently exciting current signal. Therefore, both an odd random phase multisine and a Gaussian white noise excitation were applied as an excitation signal.



\bibliography{Bibliography.bib}

\end{document}